\documentclass[12pt]{article}
\usepackage{amssymb,amsfonts,latexsym}
\usepackage{graphicx}
\usepackage{color}

\newcommand{\bbf}{{\bf f}}
\newcommand {\bR}{{\Bbb R}}

\newcommand {\bP}{{\Bbb P}}

\newcommand {\cC}{{\cal C}}
\newcommand {\cD}{{\cal D}}
\newcommand {\cF}{{\cal F}}

\newcommand {\cH}{{\cal H}}
\newcommand {\cK}{{\cal K}}
\newcommand {\cll}{{\lambda}}

\newcommand {\cO}{{\cal O}}

\newcommand{\beq}{\begin{equation}}
\newcommand{\eeq}{\end{equation}}
\newcommand{\beqn}{\begin{eqnarray}}
\newcommand{\eeqn}{\end{eqnarray}}
\newcommand{\beqno}{\begin{eqnarray*}}
\newcommand{\eeqno}{\end{eqnarray*}}
\newtheorem{theorem}{Theorem} [section]
\newtheorem{lemma}[theorem]{Lemma}
\newtheorem{propo}[theorem]{Proposition}

\newtheorem {remark}[theorem]{Remark}
\newtheorem {remarks}[theorem]{Remarks}

\begin{document}
\title {Dynamics of a classical Hall system driven by a time-dependent
  Aharonov--Bohm flux}
\author{J. Asch\thanks{CPT-CNRS,
Luminy Case 907,  F-13288 Marseille Cedex 9, France. e-mail:
asch@cpt.univ-mrs.fr }, 
  P.~\v{S}\v{t}ov\'{i}\v{c}ek\thanks{Department of Mathematics, Faculty of Nuclear Science,
  Czech Technical University, Trojanova 13, 120 00 Prague, Czech
  Republic} }

\date{04.09.2006}
\maketitle
\begin{abstract}
  We study the dynamics of a classical particle moving in a punctured
  plane under the influence of a strong homogeneous magnetic field, an
  electrical background, and driven by a time-dependent singular flux
  tube through the hole.

  We exhibit a striking classical (de)localization effect: in the  far past the trajectories are spirals around a bound center; the particle moves  inward towards the flux tube loosing kinetic energy. After hitting the puncture
 it becomes ``conducting'': the motion is a cycloid around a center whose drift is outgoing, orthogonal to the electric field, diffusive, and    without energy loss.

\end{abstract}

\section{Introduction}

The motivation to study the dynamics of this classical system is to
sharpen our intuition on its quantum counterpart which is, following
Laughlin's \cite{Laughlin} and Halperin's \cite{Halperin} proposals,
widely used for an explanation of the Integer Quantum Hall effect. Of
special interest is how the topology influences on the dynamics. In
the mathematical physics literature Bellissard et al. \cite{Bellissard} and
Avron, Seiler, Simon \cite{AvronSeilerSimonPRL},
\cite{AvronSeilerSimon} used an adiabatic limit of the model to
introduce indices. The indices explain the quantization of charge
transport observed in the experiments \cite{Klitzing}. See
\cite{CombesGerminet, ElgartGrafSchenker, CombesGerminetHislop,
  Elgart, Graf} for recent developments. We discussed the adiabatics
of the quantum system in \cite{AschHradeckyStovicek}, its quantum and
semiclassical dynamics will be treated elsewhere. The dynamics of the
classical system without magnetic field were discussed in
\cite{AschBenguriaStovicek}.

We state the model and our main results: 

Consider a classical point particle of mass $m>0$ and charge $e>0$
moving in the punctured plane $\bR^2\setminus(0)$. Suppose that a
magnetic flux line with time varying strength $\Phi$ pierces the
origin and further the presence of a homogeneous magnetic field of
strength $B>0$ orthogonal to the plane and an interior electric field
with smooth bounded potential $V$.

The equations of motions are Hamiltonian. For a point\\ $(q,p)=\left((q_{1},q_{2}),(p_{1},p_{2})\right)$ in phase space
\[\bP=\bR^2\setminus(0)\times\bR^2\]
 the time dependent Hamiltonian is :
 \[
 \frac{1}{2 m}\left(p-eA(t,q)\right)^2+eV(t,q) ; \qquad
 A(t,q)=\left(\frac{B}{2}-\frac{\Phi(t)}{2\pi\vert
     q\vert^2}\right)q^{\perp}
 \]
where $q^{\perp}:=(-q_2,q_1)$.  We suppose that 
\[\Phi:\bR\to\bR \hbox{ and } V:\bR\times\bR^2\to\bR \hbox{ are smooth functions}.\] 
The electric field is $-\partial_{t}A-\partial_{q}V$, the force on the particle with velocity $\dot q$:
\[e\left(\dot q \wedge rot(A)-\partial_{t}A-\partial_{q}V\right)=-e\left({B}\dot q^{\perp}-\frac{\partial_{t}\Phi}{2\pi}\frac{q^{\perp}}{\vert q\vert^{2}}+\partial_{q}V\right)\]

Remark that the  part of the electric field induced by the flux has
circulation $\frac{e\partial_{t}\Phi}{2\pi}$ but vanishing rotation, and is
long range with an $1/r$ singularity at the origin, we call it the circular parts. $V$ is smooth on
the entire plane so that the circulation of the corresponding field is
zero. This is the topology essential for the dynamics.
\bigskip

Recall that when only the constant magnetic field is present, the particle follows the Landau orbits; these are   circles around a fixed center with frequency $\frac{eB}{m}$ whose squared radius is proportional to the energy. 

\begin{figure}
\begin{center}
\includegraphics[width=8cm]{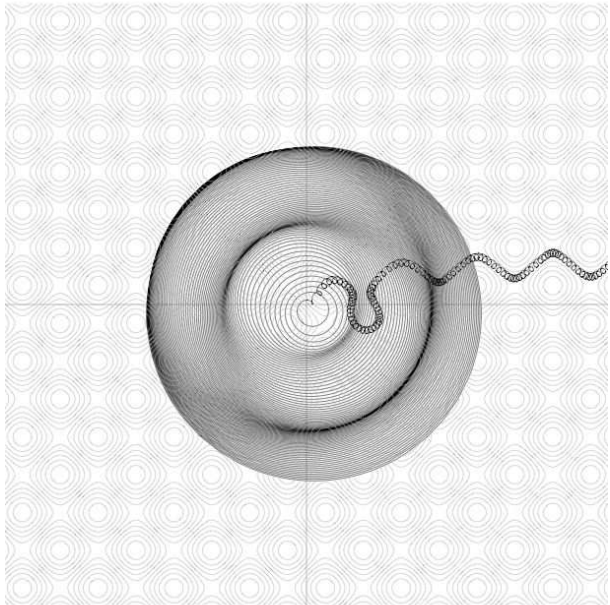}
\caption{{\small Typical trajectory of the Hamiltonian $\frac{1}{2}\left(p-\left(\frac{1}{2}q^{\perp}-s\frac{q^{\perp}}{q^{2}}+s\partial_{q}V\right)\right)^{2}$ with $V$ chosen to be $V(x, y) = 1/3 (\sin{x}+\sin{y})$}}
\label{figure1}
\end{center}
\end{figure}

Our result for the case $\Phi\sim t$, $B$ large, $V$ such that the torque $q\wedge \partial_{q}V$ is small  is qualitatively: 

\begin{itemize}
\item[{--}] the motion in configuration space is approximately
  rotation with radius proportional to the square root of the
  (time-dependent) energy around a drifting center.
\item[{--}] for large enough negative times the center is trapped by
  the flux line and the energy is linearly decreasing with time, so
  the particle is spiraling inwards
\item[{--}] from the hitting time on (i.e. the time when the Landau orbit
  ``hits'' the singularity) the center starts to drift away from the
  flux line, the energy remains asymptotically constant in the future.
  The drift is diffusive. The situation is described by
  Fig.~\ref{figure1}, showing a typical orbit in q--space.
\end{itemize}
Remark that the corresponding analysis remains true if the sign of $B$ is changed. In this case  we may state our
observation as:
{\it  Hall conducting states are eventually trapped by the
flux line and trapped states are energy conducting.}\\
Here ``hall conducting'' means that the center follows   the lines of the potential diffusively.

We shall discuss the corresponding quantum behavior elsewhere.
\bigskip

In the first section of this paper we state some general remarks on the model and discuss the problem for frozen values of the flux. Next we define appropriate action angle coordinates and use an averaging (adiabatic) method to approximate the dynamics near the hitting time between the particle and the flux line. In the last section we discuss the asymptotic behavior of the solution of the full equations of motion.

Let us remark that our method includes (for the two dimensional case) a simple proof for the guiding center approximation widely used in plasma physics.

\section{Dynamics of the frozen system}
Denote
\[\omega=\frac{eB}{m}, \cll=\frac{1}{\sqrt{eB}}.\]
We use the scaling $(t,q,p)\mapsto (\omega t, q/\cll, {p\cll})$ and ``absorb '' $V$ into the time dependent vectorpotential. The scaled variables are called $(s,q,p)$. The Hamiltonian under consideration then reads
\[H(s;p,q):=\frac{1}{2}\left(p-a(s;q)\right)^{2};\quad a(s;q):=\left(\frac{1}{2}q^{\perp}+a_{E}(s;q)\right)\]
where $a_{E}(s):\bR^{2}\setminus(0)\to\bR^{2}$ is smoothly time
dependent with $rot(a_{E})(s)=0$. $a_{E}(s)$ and the electric field
$E(s):\bR^{2}\setminus(0)\to\bR^{2}$ are defined by:
\begin{equation}
  \label{ABfield}
  -\partial_{s}a_{E}(s):=E(s)
  :=\frac{1}{\omega}\left(\frac{\partial_{t}\Phi}{2\pi}
    \!\left(\frac{s}{\omega}\right)
    \frac{q^{\perp}}{\vert q\vert^{2}}
    -\cll\, (\partial_{q} V)\!\left(\frac{s}{\omega},\cll q\right)\right)
\end{equation}

We discuss first the solution of the equation of motions for a frozen time $\sigma\in\bR$.
 As $\partial_{s}a_{E}(\sigma;q)=0$, the solution of the frozen equations  generated by the Hamiltonian $H(\sigma)$   goes along the lines of the classical Landau problem (which means: the case $\Phi=0; V(q)=0$)

\bigskip 
For  $\sigma\in\bR$ define
\begin{enumerate}
\item the velocity field: $v(\sigma ):\bP\to\bR^{2}, \quad v(\sigma ;q,p):=p-a(\sigma ;q)$;
\item the center: $c(\sigma ):\bP\to\bR^{2},\quad c(\sigma ;q,p):=q-v^{\perp}(\sigma ;q,p)$;
\item the angular momentum: $L:\bP\to\bR,\quad L(q,p):=q\wedge p$.
\end{enumerate}

Denote the Poisson bracket: $\left\lbrace f,g\right\rbrace=\partial_{q}f\partial_{p}g-\partial_{p}f\partial_{q}g$.

\bigskip
\noindent We list some useful formulas: 
\begin{propo}\label{frozenflow} The following identities hold as functions on phase space $\bP$ for all $\sigma\in\bR$:
\begin{enumerate}
\item $\left\lbrace v_{1},v_{2}\right\rbrace=1$,\quad $\left\lbrace c_{1},c_{2}\right\rbrace=-1$,\quad $\left\lbrace c,\frac{c^{2}}{2}\right\rbrace=c^{\perp}$, \quad $\left\lbrace c_{i},v_{j}\right\rbrace=0$;
\item $H=\frac{1}{2}v^{2}$,\quad $\left\lbrace v,H\right\rbrace=-v^{\perp}$,\quad $\left\lbrace c,H\right\rbrace=0$;
\item \begin{equation}
\label{centervelocityequation}
\frac{1}{2}c^{2}=\frac{1}{2}v^{2}+L- q\wedge a_{E}=H+L- q\wedge a_{E};
\end{equation}

\item the frozen flow $\left(q(\sigma ;s),p(\sigma ;s)\right)$ defined by\\ $\partial_{s}q(\sigma ;s)=\partial_{p}H(\sigma ), \quad\partial_{s}p(\sigma ;s)=-\partial_{q}H(\sigma ), \\ \quad \left(q(\sigma ;0),p(\sigma ;0)\right)=\left(q,p\right)$ is :
\begin{eqnarray*}
&&q(\sigma ;s)=c(\sigma )+\cos(s)v^{\perp}(\sigma )+\sin(s)v(\sigma )\\
&&p(\sigma ;s)=\frac{1}{2}\left(c^{\perp}(\sigma )+\cos(s)v(\sigma )-\sin(s)v^{\perp}(\sigma )\right)+a_{E}(\sigma;q(\sigma ;s))\end{eqnarray*}
\end{enumerate}
\end{propo} 

{\it Proof}: (1),(2),(3): $\left\lbrace v_{1},v_{2}\right\rbrace=\left\lbrace {p}_{1}-a_{1}(\sigma ,q), {p}_{2}-a_{2}(\sigma ,q)\right\rbrace=rot(a(\sigma ))=1$, $\left\lbrace q_{i},v_{j}\right\rbrace=\delta_{ij}$. $H=\frac{1}{2}v^{2}$ so  $\left\lbrace q,H\right\rbrace=v, \left\lbrace v,H\right\rbrace=-v^{\perp}$.
$c^{2}=q^{2}+v^{2}+2q\wedge v$; on the other hand  $L=q\wedge v+\frac{1}{2} q^{2}+ q\wedge a_{E}(\sigma;q)$.\\
(4): The force is $-\dot{q}^{\perp}$ independently of $\sigma $, Newton's
equation $\ddot{q}=-\dot{q}^{\perp}$ is readily verified. On the other
hand:
$p=v+a=a+c^{\perp}-q^{\perp}=c^{\perp}-\frac{1}{2}q^{\perp}+a_{E}(\sigma;q)$.
So $p(s)$ follows from $q(s)$ \hfill$\Box$.

\begin{remarks}
\begin{enumerate}
\item Since the energy $H(\sigma)=\frac{1}{2}v(\sigma )^{2}$ is
  conserved under the frozen flow, the projections of the trajectories
  to $q$--space are circles around $c(\sigma )$ with radius
  $\sqrt{2H(\sigma )}$. An orbit encircles the origin (has non--trivial homotopy) in
  $\bR^{2}\setminus(0)$ if and only if
\[
c^{2}<2H\Longleftrightarrow L- q\wedge a_{E}(\sigma;q)<0;
\]
\item the flow is, strictly speaking, not complete  as for $L- q\wedge a_{E}(\sigma;q)=0$ the particle  reaches the origin in $q$--space (and infinity in $p$--space) in finite time; the energy remains, however, finite. This is a mathematical subtlety which can be handled.
\end{enumerate}
\end{remarks}

\section{Action angle coordinates}

In order to discuss the full dynamics for large $B$ we introduce
action angle coordinates. The frozen dynamics as discussed in
Proposition~\ref{frozenflow} suggests to take as coordinates the angles
and absolute values of $c$ and $v^{\perp}$, i.e. with the
\[\hbox{\bf notation: }\quad e(\theta):=\left(\cos\theta, \sin\theta\right):\] 
\begin{eqnarray*}
&&q=c+v^{\perp}=\vert c\vert \frac{c}{\vert c\vert}+\vert v\vert\frac{v^{\perp}}{\vert v\vert}=:\vert c\vert e( \varphi_{1})+\vert v\vert e(- \varphi_{2})\\
&&p=\frac{1}{2}\left(c^{\perp}+v\right)+a_{E}(\sigma;q)=\frac{1}{2}\left(\vert c\vert e^{\perp}( \varphi_{1})-\vert v\vert e^{\perp}(- \varphi_{2})\right)+a_{E}(\sigma;q)
\end{eqnarray*}
Motivated by this we define  for $\sigma \in\bR$
\begin{eqnarray*}
&&q(\sigma ;\varphi,I):=\sqrt{2I_{1}}e(\varphi_{1})+\sqrt{2I_{2}}e(-\varphi_{2})\\
&&p(\sigma ;\varphi,I):=\frac{1}{2}\left(\sqrt{2I_{1}}e^{\perp}(\varphi_{1})-\sqrt{2I_{2}}e^{\perp}(-\varphi_{2})\right)+a_{E}(\sigma;q(\sigma ;\varphi,I))
\end{eqnarray*}
and, denoting by  $\cC$ the nullset $\left\{(\varphi, I); \varphi_{1}+\varphi_{2}=\pi, I_{1}=I_{2}\right\}$ where $q(\sigma ;\varphi,I)=0$, by $\cD$ the nullset $\left\{(q,p);v^{2}=0 \hbox{ or } c^{2}=0\right\}$. Thus for each frozen time $\sigma\in\bR$  the transformation to action angle coordinates $T(\sigma)$ is defined by
\begin{eqnarray*}
&&T(\sigma ):S^{1}\times S^{1}\times\left\{(I_{1},I_{2});I_{1}\ge0, I_{2}\ge0\right\}\setminus\cC\to \bP\setminus\cD\\
&&T(\sigma ;{\varphi},{I})=T(\sigma ;{\varphi_{1},\varphi_{2}},{I_{1},I_{2}}):=\left(q(\sigma ;\varphi,I),p(\sigma ;\varphi,I)\right)\\
\end{eqnarray*}

We have

\begin{lemma}
\begin{enumerate}
\item $T(\sigma )$ is a canonical diffeomorphism
\item $T^{-1}(\sigma )$ is determined by
\begin{eqnarray*}
&&I_{1}(\sigma )=\frac{c^{2}{(\sigma )}}{2}=\frac{1}{2}\left(p-\left(-\frac{1}{2}q^{\perp}+a_{E}(\sigma; q)\right)\right)^{2};\\
&& I_{2}(\sigma )=H(\sigma )=\frac{1}{2}\left(p-\left(\frac{1}{2}q^{\perp}+a_{E}(\sigma; q)\right)\right)^{2}\\
&&e(\varphi_{1}(\sigma ))=\frac{c}{\vert c\vert}(\sigma )=\frac{\frac{1}{2}q-p^{\perp}-a_{E}^{\perp}(\sigma;q)}{\sqrt{2(H(\sigma )+L-q\wedge a_{E}(\sigma;q))}}\\
&&e(-\varphi_{2}(\sigma ))=\frac{v^{\perp}}{\vert v\vert}(\sigma )=\frac{\frac{1}{2}q+p^{\perp}+a_{E}^{\perp}(\sigma; q)}{\sqrt{2H(\sigma )}}
\end{eqnarray*}
\end{enumerate}
\end{lemma}

{\it Proof:} These identities follow immediately from
Proposition~\ref{frozenflow}:
\begin{eqnarray*}
  && \{I_{1},I_{2} \}= 0, \{e(\varphi_{1}),e(\varphi_{2}) \}=0,
  \{I_{1},e(\varphi_{2})\}=0= \{I_{2},e(\varphi_{1}) \},\\
  && \{e(\varphi_{1}),I_{1} \}=\frac{1}{\vert c\vert}
  \{c,\frac{c^{2}}{2} \}
  =\frac{c^{\perp}}{\vert c\vert}=e^{\perp}(\varphi_{1}).
\end{eqnarray*}
On the other hand,
$\{e(\varphi_{1}),I_{1}\}=e^{\perp}(\varphi_{1})\{\varphi_{1},I_{1}\}$,
so $\{\varphi_{1},I_{1}\}=1$. Similarly: $\{\varphi_{2},I_{2} \}=1$.
\hfill$\Box$

\bigskip
We now investigate the full equations of motion, i.e. those for
time-dependent flux, in these action angle coordinates. As $rot(E)=0$
there exists a (possibly multi--valued) function which we denote by
$m=m(s;q)$ such that
\[
\partial_{q}m(s)=E(s)=-\partial_{s}a_{E}(s).
\]
Then $T(s)$ is generated by $m$:
\[
\partial_{s}T(s;\varphi,I)=(0,\partial_{s}a_{E}(q(s;\varphi,I))
=(\partial_{p}m,-\partial_{q}m)\circ T(s ;\varphi,I).
\]
Denote by $U(s):\bP\to\bP$ the hamiltonian flow of $H(s)$ defined by
$U(s):=(q(s),p(s))$
\[
\dot{q}(s)=\partial_{p}H,\textrm{~}\dot{p}(s)=-\partial_{q}H,
\textrm{~~}(q(0),p(0))=(q,p),
\]
then for the flow $\widehat{U}(s)=(\varphi(s),I(s))$ in action angle
coordinates defined by
\[
T(s)\circ\widehat{U}(s)=U(s)\circ T(s=0)
\]
it holds: 
\[
\dot{\varphi}(s)=\partial_{I}K\circ \widehat{U}(s),\textrm{~}
\dot{I}(s)=-\partial_{\varphi}K\circ \widehat U(s),\textrm{~~}
(\varphi(0),I(0))=(\varphi,I),
\]
where the Hamiltonian in action angle coordinates,
$K=H\circ{}T-m\circ{}T$, is
\[
K(s;\varphi, I)=I_{2}-m(s;q(s;\varphi,I))
\]
and the equations of motion are (with the notation
$\langle\cdot,\cdot\rangle$ for the scalar product)
\begin{eqnarray}
&&\dot{\varphi}(s)=\partial_{I}K=
\left(
\begin{array}{c}
0    \\
1  
\end{array}
\right) -\left\langle E(s,q(s;\varphi,I)),\partial_{I}q\right\rangle\label{phiequation} \\
&&\dot{I}(s)=-\partial_{\varphi}K=\left\langle E(s;q(s;\varphi,I)),\partial_{\varphi}q\right\rangle\label{Iequation}
\end{eqnarray}
\begin{remark}
Another way to derive these equations is to start from Newton's equation
\[\ddot q=-\dot{q}^{\perp}+E(s;q).\]
From the very definition of $c$  and $v$ one gets:
\[\dot{c}=-E^{\perp}(c+v^{\perp})\qquad \dot{v}=-v^{\perp}+E(c+v^{\perp})
\]
which in action angle coordinates gives (\ref{phiequation}), (\ref{Iequation}).
\end{remark}
\section{Averaged dynamics}
We apply  averaging with respect to the fast angle $\varphi_{2}$ to the system (\ref{phiequation}), (\ref{Iequation}) (see \cite{SandersVerhulst,Berglund}). The singularity problem can be overcome by a  regularization technique (see \cite{StiefelScheifele}). The result is that the solutions of the equations are at a distance of order $1/B$ to the solution of the averaged equations over times of  order $B$.

Remark that at this place we are mainly interested in the (de)localization effect so we did not make use of  more involved adiabatic or {\small KAM} methods in order to go to longer or even infinite time scales.

We detail this for the case 
\[
\Phi(t)=\Phi_{0}t,\quad V \hbox{ time independent,}
\]
i.e., a flux $\Phi_{0}$ per unit time is added ad-eternam.

Denote the average of a function $f$ on the phase space by
\[
f_{av}(\varphi_{1},I) := \frac{1}{2\pi}\int_{0}^{2\pi}
f(\varphi_{1},\varphi_{2},I)\ d\varphi_{2}
\]
In particular for a function $f$  defined on the plane thus depending  only on the variable $q$ we denote
\[
f_{av}(\varphi_{1},I)
= \frac{1}{2\pi}\int_{0}^{2\pi}f\!\left(\sqrt{2I_{1}}e(\varphi_{1})
  +\sqrt{2I_{2}}e(-\varphi_{2})\right)\ d\varphi_{2}
\]
 
 The field  (\ref{ABfield}) is
 \[E(s;q)=\frac{e}{\omega}\left(\frac{\Phi_{0}}{2\pi}\frac{q^{\perp}}{\vert q\vert^{2}}-\cll (\partial_{q}V)\left(\cll q\right)\right)\]
 Define
\[{\bbf}:=\frac{e\Phi_{0}}{2\pi\omega}\]
and choose $m$ and thus $K$:
 \begin{eqnarray*}
   m(q) &=& \bbf\  arg(q)- \frac{e}{\omega}\,V\!\left(\cll q\right)\\
   K(\varphi,I) &=& I_{2}-m\!\left(\sqrt{2I_{1}}e(\varphi_{1})
     +\sqrt{2I_{2}}e(-\varphi_{2})\right)\\
\end{eqnarray*}

Making use of the identities
\[\left\langle \frac{q^{\perp}}{q^{2}},\partial_{I}q\right\rangle=\frac{\sin(\varphi_{1}+\varphi_{2})}{q^{2}}
\left(
\begin{array}{c}
\sqrt{\frac{I_{2}}{I_{1}}} \\
-\sqrt{\frac{I_{1}}{I_{2}}}  
\end{array}\right),\quad
\left\langle  \frac{q^{\perp}}{q^{2}},\partial_{\varphi}q\right\rangle=
\left(
\begin{array}{c}
\frac{I_{1}-I_{2}}{q^{2}}+\frac{1}{2}\\
\frac{I_{1}-I_{2}}{q^{2}}-\frac{1}{2}
\end{array}\right)
\]
the system (\ref{phiequation}), (\ref{Iequation}) reads

\begin{eqnarray*}
&&\dot{\varphi}(s)=
\left(
\begin{array}{c}
0    \\
1  
\end{array}
\right) -\bbf \frac{\sin(\varphi_{1}+\varphi_{2})}{2\left(I_{1}+I_{2}+2\sqrt{I_{1}I_{2}}\cos(\varphi_{1}+
\varphi_{2})\right)}
\left(
\begin{array}{c}
\sqrt{\frac{I_{2}}{I_{1}}} \\
-\sqrt{\frac{I_{1}}{I_{2}}}  
\end{array}\right) +\frac{e}{\omega}\,\partial_{I}V(\lambda q)\\
&&\dot{I}(s)=
\bbf\frac{I_{1}-I_{2}}{2\left(I_{1}+I_{2}+2\sqrt{I_{1}I_{2}}
\cos(\varphi_{1}+\varphi_{2})\right)}\left(
\begin{array}{c}
1\\
1\end{array}\right)+\frac{\bbf}{2}\left(
\begin{array}{r}
1\\
-1
\end{array}\right)-\frac{e}{\omega}\,\partial_{\varphi}V(\lambda q)
\end{eqnarray*}

The averaged quantities are readily calculated: using 
\[
\left(\frac{1}{q^{2}}\right)_{av}
=\frac{1}{2\vert I_{1}-I_{2}\vert}\,,\qquad
\left(\frac{\sin{(\varphi_{1}+\varphi_{2}})}{q^{2}}\right)_{av}=0,
\]
one finds for the averaged vectorfield

\begin{eqnarray}
(\partial_{I}K)_{av}(\varphi_1,I) &=& \left(
\begin{array}{c}
0 \\
1
\end{array}\right)
+ \frac{e}{\omega}\,\partial_{I}V_{av}(\varphi_1,\cll^2I)
\label{phiequationav}\\
-(\partial_{\varphi}K)_{av}(\varphi_1,I) &=& \bbf\left(
\begin{array}{c}
\chi(I_{1}>I_{2}) \\
-\chi(I_{1}<I_{2})
\end{array}\right)-
\frac{e}{\omega}\left(\begin{array}{c}
\partial_{\varphi_{1}}V_{av}(\varphi_1,\cll^2I) \\
0
\end{array}
\right)\nonumber
\end{eqnarray}
where we used the binary function $\chi$:
$\chi(True):=1,\quad\chi(False):=0$.  

\begin{remark}
  Remark that the averaged vectorfield is the hamiltonian vectorfield derived from the
  from the ``averaged'' Hamiltonian $K_{av}$. Indeed, using the splitting of  $arg(q)$, which is a  multi-valued function defined on the covering space of $\bR^{2}\setminus(0)$,    into a linear and oscillating part
\begin{eqnarray*}
\arg\left(q(\varphi,I)\right) & = & \left\{ \begin{array}{cc}
\varphi_{1}+\arg\!\left((1,0)+\sqrt{\frac{I_{2}}{I_{1}}}\, e(-\varphi_{1}-\varphi_{2})\right) & \textrm{if }I_{1}>I_{2}\\
\\-\varphi_{2}+\arg\!\left((1,0)+\sqrt{\frac{I_{1}}{I_{2}}}\, e(\varphi_{1}+\varphi_{2})\right) & \textrm{if }I_{2}>I_{1}\end{array}\right..\end{eqnarray*}
and:
\[\int_{0}^{2\pi}\arg((1,0)+a\, e( s))\, ds=0\quad{\rm for }\quad  0\leq a<1; \]
One finds that for
\begin{displaymath}
  K_{av}(\varphi, I):=I_{2}
  -\frac{e}{\omega}\left(\frac{\Phi_{0}}{2\pi}
    \Big(\varphi_{1}\,\chi(I_{1}> I_{2})
    - \varphi_{2}\,\chi(I_{1}< I_{2})\Big)
    - V_{av}(\varphi_1,\cll^2I)\right)\\
\end{displaymath}
one has
$\partial_\varphi{}K_{av}=(\partial_\varphi{}K)_{av}$,
$\partial_I{}K_{av}=(\partial_I{}K)_{av}$.

\end{remark}

The result on the dynamics now is:

\begin{theorem} Denote by $J=(J_{1},J_{2}), \quad \psi=(\psi_{1},\psi_{2})$ the solution of the averaged equations (\ref{phiequationav})
\begin{eqnarray*}
\dot{\psi}(s)&=&\partial_{I}K_{av}(\psi(s), J(s)),\quad J(0)=(J_{1}^{0},J_{2}^{0})\\
\dot{J}(s)&=&-\partial_{\varphi}K_{av}(\psi(s), J(s)),\quad \psi(0)=(\psi_{1}^{0},\psi_{2}^{0})
\end{eqnarray*}
and  by $I=(I_{1},I_{2}), \quad \varphi=(\varphi_{1},\varphi_{2})$ the solution of the full equations (\ref{phiequation}), (\ref{Iequation})
\begin{eqnarray*}
\dot{\varphi}(s)&=&\partial_{I}K(\varphi(s), I(s)),\quad I(0)=(I_{1}^{0},I_{2}^{0})\\
\dot{I}(s)&=&-\partial_{\varphi}K(\varphi(s), I(s)),\quad \varphi(0)=(\varphi_{1}^{0},\varphi_{2}^{0})
\end{eqnarray*}
then it holds

\begin{enumerate}
\item Let $V=0$, denote  $\Delta J=J_{2}^{0}-J_{1}^{0}$ then:
\begin{eqnarray*}
&&J(s)=\min\{J_{1}^{0},J_{2}^{0}\}+(\bbf s-\Delta J)
\left(\begin{array}{r}
\chi\left(\bbf s>\Delta J\right) \\
-\chi\left(\bbf s<\Delta J\right)\end{array}\right)
\\
&&\psi(s)=
\left(\begin{array}{c}
\psi_{1}^{0} \\
\psi_{2}^{0}+s\end{array}\right)
\end{eqnarray*}
\item For any $V$ and any $s_{1},s_{2}\in\bR$
\[\vert J_{2}(s_{2})-J_{2}(s_{1})\vert=\bbf\left\vert\int_{s_{1}}^{s_{2}}\chi\left(J_{1}(u)<J_{2}(u)\right)du \right\vert\]
\item Let $V$ be such that  the torque of the corresponding field satisfies for a $c\in\lbrack0,1)$:
\[\vert q\wedge\partial_{q}V\vert\le \frac{\Phi_{0}}{2\pi} c\]
then for any initial condition it holds:
\[I_{1}-I_{2}\hbox{ is strictly increasing, furthermore}\]
\[\bbf(1-c)\le \dot{I_{1}}(s)-\dot{I_{2}}(s)\le \bbf(1+c)\qquad(\forall s\in\bR).\]
\item In particular if $q\wedge\partial_{q}V=0$ it holds for all $s\in\bR$:
\begin{equation}\label{I1-I2}
{I_{1}}(s)-{I_{2}}(s)=\bbf (s-s_{0})
\end{equation}

where $s_{0}$ is the unique ``hitting time'' defined by this equation.
\end{enumerate}
\end{theorem}

{\it Proof}: Using that for $V=0$ it holds
$J_{1}(s)-J_{2}(s)-\bbf{}s=\Delta{}J$ the first assertion follows by
inspection. The second assertion follows from integration of
(\ref{phiequationav}). Finally we have from
(\ref{centervelocityequation}):
\begin{eqnarray*}
&&\dot{I_{1}}-\dot{I_{2}}=\partial_{s}(I_{1}-I_{2})=q\wedge E=\frac{e}{\omega}\left(\frac{\Phi_{0}}{2\pi}-(q\wedge \partial_{q}V)(\lambda q)\right)\\
\end{eqnarray*}
from which the last assertion follows.
\hfill$\Box$

\begin{remarks}
\begin{enumerate}
\item The first equation explains the qualitative behavior of the
  solution exhibited in Fig.~\ref{figure1}: $J_{1}$ is linear in time
  in the future and is constant in the past.
\item Loosely speaking the second assertion of the theorem means that, on the average, one has
\[
\vert\hbox{energychange}\vert
=\vert\hbox{fluxchange through the orbit during stay time}\vert
\]
where the stay time means the time where the ``orbit surrounds the
origin''. This should be like this as the the change in energy equals
the work of the electric field along the orbit:
\[H(s;q(s))-H(s_{0};q(s_{0}))=\int_{s_{0}}^{s}\langle a_{E}(s),ds\rangle.\]
\item The last assertion says that the orbit presented in
  Fig.~\ref{figure1} in the introduction is generic, i.e.: inward
  spiraling motion with fixed center followed by the usual Hall
  cycloids with the center following the lines of the potential. We
  argue that our condition on the potential is far from optimal and
  that for large enough magnetic field the situation described in this
  paper is generic for $V$ smooth and bounded with bounded
  derivative. This needs further investigation.
\end{enumerate}

\end{remarks}

\section{Large time asymptotics, potential free case}

For the case $\Phi(t)=\Phi_{0}t$, $V=0$ we can determine the large
time asymptotics of the solution. We keep the notation
${\bbf}:=\frac{e\Phi_{0}}{2\pi\omega}$. Observe also that
\begin{displaymath}
  K = K(\varphi,I) =
  I_2-\arg\!\left(\sqrt{2I_1}\,e(\varphi_1)+\sqrt{2I_2}\,e(-\varphi_2)\right)
\end{displaymath}
is an integral of motion.

\begin{theorem}
  Denote by $I=(I_{1},I_{2})$, $\varphi=(\varphi_{1},\varphi_{2})$ the
  solution of the full equations of motion (\ref{phiequation}),
  (\ref{Iequation})
  \begin{eqnarray*}
    \dot{\varphi}(s) &=& \partial_{I}K(\varphi(s), I(s)),\quad
    I(0)=(I_{1}^{0},I_{2}^{0})\\
    \dot{I}(s) &=& -\partial_{\varphi}K(\varphi(s), I(s)),\quad
    \varphi(0)=(\varphi_{1}^{0},\varphi_{2}^{0})
  \end{eqnarray*}
  then the following asymptotic behavior holds:

  \underline{in the future, $s\to\infty$}

  The following limits exist and define the constants $a_{0}>0$,
  $b_{0}$:
  \[
  \lim_{s\to\infty}I_{2}(s)=:\frac{a_{0}^{2}}{4\bbf}\,,\quad
  \lim_{s\to\infty}\left(\varphi_{1}(s)+\varphi_{2}(s)-s\right)=:b_{0},\quad
  \lim_{s\to\infty}\left(I_{2}(s)-\bbf \varphi_{1}(s)\right)=K,
  \]
  the asymptotics are
  \begin{eqnarray*}
    I_{2}(s) &=& \frac{a_{0}^{2}}{4\bbf}
    -\left(\frac{a_{0}}{2}\sin(s+b_{0})\right)
    \frac{1}{\sqrt s}+\frac{1}{4}\left(\bbf+\frac{a_{0}^{2}}{2\bbf}
      \sin(2(s+b_{0}))\right)\frac{1}{s}
    +\cO\!\left(\frac{1}{s^{3/2}}\right)\\
    I_{1}(s) &=& I_{2}(s)+\bbf(s-s_{0})\\
    \varphi_{1}(s) &=& \frac{a_{0}^{2}}{4\bbf^{2}}-\frac{K}{\bbf}
    -\frac{1}{4s}+\cO\!\left(\frac{1}{s^{3/2}}\right)\\
    \varphi_{2}(s) &=& s+b_{0}-\frac{a_{0}^{2}}{4\bbf^{2}}
    +\frac{K}{\bbf}-\frac{\bbf}{a_{0}}\cos(s+b_{0})\frac{1}{\sqrt{s}}\\
    && +\,\frac{1}{8}\left(-1+2\cos(2(s+b_{0}))
      -\frac{4\bbf^{2}}{a_{0}^{2}}\sin(2(s+b_{0}))\right)\frac{1}{s}
    +\cO\!\left(\frac{1}{s^{3/2}}\right)
  \end{eqnarray*}
  with $s_{0}$ defined as in (\ref{I1-I2});

  \bigskip
  \underline{in the past, $s\to-\infty$}

  The following limits exist and define the constants
  $\widetilde{a}_{0}>0$, $\widetilde{b}_{0}$:
  \[
  \lim_{s\to-\infty}I_{1}(s)=:\frac{\widetilde a_{0}^{2}}{4\bbf}\,,\quad
  \lim_{s\to-\infty}\left(\varphi_{1}(s)+\varphi_{2}(s)-s\right)=:
  \widetilde b_{0},\quad
  \lim_{s\to-\infty}\left(I_{2}(s)+\bbf\varphi_{2}(s)\right)=K,
  \]
  the asymptotics are
  \begin{eqnarray*}
    I_{1}(s) &=& \frac{\widetilde a_{0}^{2}}{4\bbf}
    +\left(\frac{\widetilde a_{0}}{2}\sin(s+\widetilde b_{0})\right)
    \frac{1}{\sqrt{\vert s\vert}}
    -\frac{1}{4}\left(\bbf-\frac{\widetilde a_{0}^{2}}{2\bbf}
      \sin(2(s+\widetilde b_{0}))\right)\frac{1}{s}
    +\cO\!\left(\frac{1}{\vert s\vert^{3/2}}\right)\\
    I_{2}(s) &=& I_{1}(s)-\bbf(s-s_{0})\\
\noalign{\vspace{0.1\baselineskip}}
    \varphi_{1}(s) &=& s_{0}+\widetilde b_{0}
    +\frac{\widetilde a_{0}^{2}}{4\bbf^{2}}-\frac{K}{\bbf}
    +\frac{\bbf}{\widetilde a_{0}}\cos(s+\widetilde b_{0})
    \frac{1}{\sqrt{\vert s\vert}}\\
    && -\,\frac{1}{8}\left(1-2\cos(2(s+\widetilde b_{0}))
      -\frac{4\bbf^{2}}{\widetilde a_{0}^{2}}
      \sin(2(s+\widetilde b_{0}))\right)\frac{1}{s}
    +\cO\!\left(\frac{1}{\vert s\vert^{3/2}}\right)\\
    \varphi_{2}(s) &=& s-s_{0}
    -\frac{\widetilde a_{0}^{2}}{4\bbf^{2}}
    +\frac{K}{\bbf}-\frac{1}{4s}
    +\cO\!\left(\frac{1}{\vert s\vert^{3/2}}\right).
  \end{eqnarray*}
\end{theorem}

{\it Proof}: We give an outline of the main steps of the proof for the
case $t\to\infty$. Some particular computations in the proof turned
out to be quite tedious and thus computer algebra systems were
employed to facilitate them.

Suppose $t>0$

\bigskip
\underline{Step 1}

From (\ref{I1-I2}) we know $I_{1}(s)-I_{2}(s)=\bbf(s-s_{0})$. So the
equations of motion only involve $J:=I_{1}+I_{2}$ and
$\psi:=\varphi_{1}+\varphi_{2}$ and transform to
\[
\dot\psi = 1+\frac{\bbf^{2}s \sin\psi}{\sqrt{J^{2}-\bbf^{2}s^{2}}
  (J+\sqrt{J^{2}-\bbf^{2}s^{2}}\cos\psi)}\,,\textrm{~~}
\dot J=\frac{\bbf^{2}s}{J+\sqrt{J^{2}-\bbf^{2}s^{2}}\cos\psi}\,,
\]

\underline{Step 2}

Do a second transformation
\[
x_{1}:=\sqrt{J^{2}-\bbf^{2}s^{2}}\cos\psi,\textrm{~~}
x_{2}:=\sqrt{J^{2}-\bbf^{2}s^{2}}\sin\psi,
\]
the $J, \psi$ equations transform to 
\begin{displaymath}
\dot{x}_{1}-\frac{x_{1}}{s}+x_{2}=F(s,x_{1},x_{2}),\textrm{~~}
\dot{x}_{2}-x_{1}=0,
\end{displaymath}
with 
\[
F(s,x_{1},x_{2}) := \bbf -\frac{x_{1}}{s}
-\frac{\bbf^{2}s}{\sqrt{x_{1}^{2}+(x_{2}-\bbf)^{2}+\bbf^{2}s^{2}}+x_{1}}\,.
\]

The corresponding homogeneous system is equivalent to 
\begin{eqnarray*}
\ddot{x}_{1}-\frac{\dot{x}_{1}}{s}+\left(1+\frac{1}{s^{2}}\right)x_{1}=0
\textrm{~~~or~~~}
s\ddot{y}+\dot{y}+sy = 0
\end{eqnarray*}
with $y$ defined by $x_{1}=sy$. The latter is Bessel's equation of order $0$ so one has two independent solutions of the homogeneous system:
\begin{eqnarray*}
\left(
\begin{array}{c}
 x_{1}(s)  \\
x_{2} (s)
\end{array}
\right)=
\left(
\begin{array}{c}
sJ_{0}(s) \\
sJ_{1}(s)
\end{array}
\right)\quad \hbox{\rm and } \quad
\left(
\begin{array}{c}
 x_{1}(s)  \\
x_{2} (s)
\end{array}
\right)=
\left(
\begin{array}{c}
sY_{0}(s) \\
sY_{1}(s)
\end{array}
\right)
\end{eqnarray*}
with the Bessel functions $J_{m}$ ($Y_{m}$) of the first (second) kind.

\bigskip
\underline{Step 3}

Transform the $x$--differential equation to the integral equation
\begin{eqnarray*}
  x_{1}(s) &=& c_{1}sJ_{0}(s)+c_{2}sY_{0}(s)\\
  && -\,\frac{\pi s}{2}\int_{s}^{\infty}
  (Y_{0}(s)J_{1}(\tau)-J_{0}(s)Y_{1}(\tau))
  F(\tau,x_{1}(\tau),x_{2}(\tau))d\tau\\
  x_{2}(s) &=& c_{1}sJ_{1}(s)+c_{2}sY_{1}(s)\\
  && -\,\frac{\pi s}{2}\int_{s}^{\infty}
  (Y_{1}(s)J_{1}(\tau)-J_{1}(s)Y_{1}(\tau))
  F(\tau,x_{1}(\tau),x_{2}(\tau))d\tau
\end{eqnarray*}
where the numbers $c_{1}, c_{2}$ involve the initial conditions.

The equation is of the form $x=\cK(x)$, the solution is constructed as
the limit of the sequence $x_{n+1}=\cK(x_{n})$ starting from
$x_{0}=0$. To verify the convergence one can apply yet another
substitution $x(s)=y(s)/\sqrt{s}$,
$G(s,y)=s^{-1/2}F(s,s^{-1/2}y)$. Consequently the integral equation
takes the form
\begin{displaymath}
  y(s) = y_{0}(s)-\int_{s}^{\infty}\cF(s,\tau)\,
  G(\tau,y_{1}(\tau),y_{2}(\tau))\,d\tau
\end{displaymath}
where
\begin{eqnarray*}
  y_{0j}(s) &=& c_{1}\sqrt{s}\,J_{j-1}(s)+c_{2}\sqrt{s}\,Y_{j-1}(s),
  \textrm{~~}j=1,2,\\
  \cF_{j}(s,\tau) &=& \frac{\pi}{2}\,\sqrt{s\tau}\,
  \big(Y_{j-1}(s)J_{1}(\tau)-J_{j-1}(s)Y_{1}(\tau)\big),
  \textrm{~~}j=1,2.
\end{eqnarray*}
Considering the new integral equation in the Banach space
$L^\infty([\,s_\ast,\infty[)\otimes\bR^2$ one can show that the
iteration process is indeed contracting provided $s_\ast\geq1$ is
sufficiently large. It is then straightforward to derive from the
integral equation the asymptotic expansion of the solution $x(s)$. One
finds that
\[
x(s) = a_{0}\,e(t+b_{0})\sqrt{s}
+\left(\frac{a_{0}^{3}}{8\bbf^{2}}\,e(t+b_{0})
  -\frac{5}{8}a_{0}\,e^{\perp}(t+b_{0})\right)
\frac{1}{\sqrt{s}}+\cO\!\left(\frac{1}{s}\right)
\]

\underline{Step 4}

Transforming back first to the $J,\psi$ then to
$I_{1},I_{2},\varphi_1,\varphi_2$ variables gives the claimed
asymptotic expansion.  \hfill$\Box$

The asymptotic formulae for the actions and the angles imply the
following asymptotic behavior of the solutions and the energy thus
defining the transport coefficients:

Denote 
\[\cH:= 
 \frac{1}{2 m}\left(p-e\left(\frac{B}{2}-\frac{\Phi(t)}{2\pi\vert
    q\vert^2}\right)q^{\perp}\right)^2
\]
the energy in the original coordinates $q,p$, and $q_{sc}=q/\lambda$
the scaled coordinate. Rescaling then gives
\[
\cH(t)=\omega H(\omega t)=\omega I_{2}(\omega t),\textrm{~}
q(t)=\lambda q_{sc}(\omega t),\textrm{~}
q_{sc}=\sqrt{2I_{1}}e(\varphi_{1})+\sqrt{2I_{2}}e(-\varphi_{2}).
\]
This leads to the following limits valid for any fixed initial
condition and any $B>0,\Phi_{0}>0$:
\begin{eqnarray*}
  \frac{q(t)}{\sqrt{t}} &\to_{t\to\infty}& 
  \sqrt{\frac{\Phi_{0}}{2\pi B}}\,
  e\!\left(\frac{a_{0}^{2}}{4\bbf^2}-\frac{K}{\bbf}\right)\\
  \frac{q(t)}{\sqrt{|t|}} &\sim_{t\to-\infty}&
  \sqrt{\frac{\Phi_{0}}{2\pi B}}\,e(-\omega t)\\
  \cH(t) &\to_{t\to\infty}& \frac{\omega a_{0}^{2}}{4\bbf}\\
  \frac{\cH(t)}{t} &\to_{t\to-\infty}&
  -\frac{e^{2}B}{m}\,\frac{\dot{\Phi}}{2\pi} = -\frac{e^{2}B}{m}\frac{\Phi_{0}}{2\pi}
\end{eqnarray*}

\section*{Acknowledgments}
P.~\v{S}. wishes to acknowledge gratefully partial support from the
grants\\ No.~201/05/0857 of the Grant Agency of the Czech Republic and
No.~LC06002 of the Ministry of Education of the Czech Republic.


\begin{thebibliography}{1}

  \bibitem{AschBenguriaStovicek}
Asch, J. and Benguria, R. D. and \v{S}\v{t}ov\'\i\v{c}ek, P.,
     ``Asymptotic properties of the differential equation {$h\sp
              3(h''+h')=1$},'' {Asymptot. Anal.}  \textbf{41}, 23--40 (2005).
  
  \bibitem{AschHradeckyStovicek}Asch, J. and Hradeck{\'y}, I. and
              \v{S}\v{t}ov\'\i\v{c}ek, P., ``Propagators weakly associated to a family of {H}amiltonians
              and the adiabatic theorem for the {L}andau {H}amiltonian with
              a time-dependent {A}haronov-{B}ohm flux,''
J. Math. Phys. \textbf{46}, {053303 ff.} (2005).


\bibitem{AvronSeilerSimonPRL}Avron, J.~E., Seiler, R., and Simon, B.,
  ``Quantum Hall Effect and the Relative Index for Projections,''
  Phys. Rev. Lett. \textbf{65}, 2185-2188 (1990).
  
\bibitem{AvronSeilerSimon}Avron, J.~E., Seiler, R., and Simon, B.,
  ``Charge deficiency, charge transport and comparison of
    dimensions,''
  Commun. Math. Phys. \textbf{159}, 399-422 (1994).

\bibitem{Bellissard}Bellissard, J., van Elst, A., and Schulz-Baldes, H.,
  ``The noncommutative geometry of the quantum Hall effect,''
  J. Math. Phys. \textbf{35}, 5373-5451 (1994).
  
\bibitem{Berglund}Berglund, N.
  ``Perturbation theory of dynamical systems,''
 Lecture Notes. http://arXiv.org/abs/math.HO/0111178 (2001).

\bibitem{CombesGerminet}Combes, J.-M., and Germinet, F., ``Edge and impurity effects on quantization of Hall 
currents,'' Comm. Math. Phys. \textbf{256}, 159-180, (2005)

\bibitem{CombesGerminetHislop}
Combes, J.-M.,  Germinet, F., and Hislop, P.D.
\newblock On the Quantization of Hall Currents in Presence of Disorder.
\newblock In Asch, J. and Joye, A. (eds), \emph{Mathematical Physics of Quantum Mechanics}, Lecture Notes in Physics , Vol. 690, (Springer, New York, 2006).

\bibitem{Elgart}
Elgart, A.,
\newblock Equality of the Bulk and Edge Hall Conductances in 2D.
\newblock In Asch, J. and Joye, A. (eds), \emph{Mathematical Physics of Quantum Mechanics}, Lecture Notes in Physics , Vol. 690, (Springer, New York, 2006).
  

\bibitem{ElgartGrafSchenker}Elgart, A., and Graf, G. M.,  and
  Schenker, J. H.,  ``Equality of the bulk and edge Hall 
conductances in a mobility gap,'' Comm. Math. Phys. \textbf{259},185Ð221, (2005). 

\bibitem{Graf} Graf, G. M.,  ``Aspects of the integer quantum Hall effect,'' Simon Festschrift,  (2006). 

\bibitem{Halperin}Halperin, B.~I., ``Quantized Hall Conductance,
    Current-Carrying Edge States and the Existence of Extended States
    in a Two-Dimensional Disordered Potential,''
  Phys. Rev. B \textbf{25}, 2185-2188 (1982).
  
\bibitem{Klitzing}von~Klitzing, K., Dorda, G., and Pepper, M., ``New
    method for high-accuracy determination of the fine-structure
    constant based on quantized hall resistance,''
  Phys. Rev. Lett. \textbf{45}, 494-497 (1980).

\bibitem{Laughlin}Laughlin, R.~B., ``Quantized Hall conductivity in
    two dimensions,''
  Phys. Rev.~B \textbf{23}, 5632-5633 (1981).

\bibitem{SandersVerhulst}J. A. Sanders\ and\ F. Verhulst, \emph{Averaging methods in nonlinear dynamical systems}, (Springer, New York, 1985)

\bibitem{StiefelScheifele}E. L. Stiefel\ and\ G. Scheifele, \emph{Linear and regular celestial mechanics. Perturbed two-body motion, numerical methods, canonical theory}, (Springer, New York, 1971)

\end{thebibliography}
\end{document}